\documentclass[useAMS,usenatbib]{mn2e}
\usepackage{epsf}

\def\Alfven{Alfv\'{e}n~}
\def\alpp{\alpha_{\rm sh,\perp}}
\def\alpr{\alpha_{\rm sh,\parallel}}
\def\Fwpp{F_{\rm w,\perp}}
\def\Fwppi{F_{\rm w,\perp,0}}
\def\Fwpr{F_{\rm w,\parallel}}
\def\Fwpri{F_{\rm w,\parallel,0}}
\def\Spp{S_{\rm w,\perp}}
\def\fmax{f_{\rm max}}
\def\mbf#1{\mbox{\boldmath ${#1}$}}
\def\lesssim{\; \buildrel < \over \sim \;}
\def\gtrsim{\; \buildrel > \over \sim \;}

\def\apj{ApJ}

\def\apjl{ApJL}
\def\aap{A\&A}
\def\aaps{A\&AS}
\def\jgr{JGR}
\def\ssr{Space Science Review}
\def\prl{Phys.Rev.Lett.}

\def\soph{Sol.Phys.}

\title[Coronal Heating and Solar Wind Acceleration by MHD shock trains]
{Coronal Heating and Acceleration of the High/Low-Speed Solar Wind 
by Fast/Slow MHD Shock Trains}
\author[T. K. Suzuki]{Takeru K. Suzuki$^{1}$\thanks{E-mail:
stakeru@scphys.kyoto-u.ac.jp}\\
$^{1}$ Departiment of Physics, Kyoto University, Kyoto 606-8502, Japan; 
JSPS Research Fellow}
\begin{document}

%\date{Accepted 2003 December 19. Received 2003 November 28; in original form 2003 Septemper 23}
\date{MNRAS, in press}
\pagerange{\pageref{firstpage}--\pageref{lastpage}} \pubyear{2002}

\maketitle

\label{firstpage}

\begin{abstract}
We investigate coronal heating and acceleration of the high- and low-speed 
solar wind in the open field region by dissipation of fast and slow 
magnetohydrodynamical (MHD) waves through MHD shocks. 
Linearly polarized \Alfven (fast MHD) waves and 
acoustic (slow MHD) waves travelling upwardly along with a magnetic 
field line eventually form fast switch-on shock trains and hydrodynamical 
shock trains (N-waves) respectively to heat and accelerate the plasma. 
We determine one dimensional structure of the corona from 
the bottom of the transition region (TR) to 1AU under the 
steady-state condition by solving 
evolutionary equations for the shock amplitudes simultaneously with 
the momentum and proton/electron energy equations.
Our model reproduces the overall trend of the high-speed wind from the polar 
holes and the low-speed wind from the mid- to low-latitude streamer 
except the observed hot corona in the streamer.
The heating from the slow waves is effective in the low corona to increase the 
density there, and plays an important role in the formation of the dense 
low-speed wind. 
On the other hand, the fast waves can carry a sizable 
energy to the upper level to heat the outer corona and accelerate 
the high-speed wind effectively.
We also study dependency on field strength, $B_0$, at the bottom of the 
TR and non-radial expansion of a flow tube, $f_{\rm max}$, to find that  
large $B_0/f_{\rm max}\gtrsim 2$ but small $B_0\simeq 2$G are favorable for  
the high-speed wind and that small $B_0/f_{\rm max}\simeq 1$ is 
required for the low-speed wind. 

\end{abstract}


\section{Introduction}
The problems of the coronal heating and the solar wind acceleration should 
be investigated simultaneously 
since they are very closely related, although they have been often 
discussed separately. 
Even combined models, taking into account them together, usually adopt too 
simplified heating and acceleration law:  
they adopt unspecified mechanical energy and momentum 
inputs of the exponential type, 
$\propto \exp(\frac{-(r-R_{\odot})}{l_{\rm m}})$, with an assumed constant 
dissipation length, $l_{\rm m}$, for lack of an alternative 
(e.g.Sandb{\ae}k \& Leer 1994), although the constant $l_{\rm m}$ is poorly 
supported by fundamental physical processes.
Therefore, it is required to treat the energy transfer and momentum transfer 
simultaneously in the 
broad corona including the solar wind with realistic heating and acceleration 
functions. 

The origin of the energy to heat the corona and accelerate the solar wind 
is generally believed to lie in the turbulent convective motions beneath 
the photosphere. 
The major difficulties in understanding the problems are how to lift up the 
energy to the corona and let it dissipate at the appropriate locations. 
Various modes of waves are supposed to play important roles in these 
processes (see Roberts 2000 for recent review).
 Not only the surface granulations but also dynamical activities 
such as the small magnetic reconnection events 
\citep{tsu96,trc99,sky00,kt01} excites the waves from 
at the photospheric level to in the corona \citep{str99}. 
MHD waves have been studies long as reliable heating and acceleration 
sources of the solar coronal plasma
(e.g. Barnes 1969; Belcher 1971). 
Many groups have examined advanced properties on the wave propagation and 
damping such as non-WKB effects \citep{mc94,od98}, steepening of slow 
\citep{ons00} and fast \citep{noa00} MHD waves, and shock formation 
\citep{hol92} to study the wave heating.   
As an effect beyond the MHD, resonant dissipation of 
ion-cyclotron waves \citep{dh81,mcr82} has also been highlighted as a heating 
source of the heavy ions in the high-speed solar wind \citep{cfk99,hh99}. 
However, because of the too high resonant frequencies, the protons and 
electrons, underlying the coronal plasma, are expected to be mainly heated by 
the MHD waves with lower frequency in which the most of the wave energy 
is contained. 
Signatures of 
such MHD waves have been observed in various portions of the solar corona 
\citep{hrsh90,btd98,ond99,omf02}, whereas dominant processes in the 
wave heating yet remain to be determined.

When one considers the MHD waves travelling along with a magnetic field line 
in the low-$\beta$ plasma, they can be divided in two types in the linear 
regime where the wave amplitude is sufficiently smaller than the phase speed. 
One is slow-mode wave which is longitudinal wave and essentially 
identical to the acoustic wave. The other is \Alfven wave which is
transverse wave. When the non-linear terms are taken into account, the 
degeneracy of the transverse waves is removed; they can be divided into 
circularly polarized \Alfven wave classified as the MHD intermediate-mode, 
and linearly 
polarized \Alfven wave classified as the MHD fast-mode.   
Among these three types of the MHD waves, 
the fast-mode and slow-mode waves nonlinearly steepen their shapes through 
propagation to eventually form the MHD fast shocks and slow shocks 
respectively.  
Suzuki (2002; hereafter S02) investigated the dissipation of the N-wave 
which is a special case of the slow-shock trains 
propagating along the field line. 
S02 also inspected the consequent heating in self-consistent 
modelling to conclude that the N-waves can heat the inner corona effectively 
though they cannot heat the entire corona. 
The heating by the dissipation of the switch-on shock trains, a special case 
of the fast shock trains, was proposed by Hollweg (1982; hereafter H82). 
He studied the problem on fixed 
coronal structures to conclude that it can be a reliable process of 
the coronal heating. However, the self-consistent treatment of their 
propagation and dissipation has not been performed so far.    

In this paper, we investigate the coronal heating and the solar wind 
acceleration by the switch-on shock trains as well as the N-waves 
in a self-consistent manner. First, by extending the formulation 
introduced by H82, we derive an equation describing variation of the 
switch-on shock trains in the similar way to that adopted by \citet{ss72} 
and S02 for the N-waves. 
We solve these equations for the shock trains simultaneously 
with the equations of the momentum transfer and energy transfer to 
determine the unique structure of the corona and solar wind from the bottom of 
the transition region (TR) to 1AU for given wave energy flux of the linearly 
polarized \Alfven waves, $\Fwppi$, and acoustic waves, $\Fwpri$, at the 
coronal base.
The heating and the acceleration by 
the dissipation of the shock trains are explicitly taken into account. 
Therefore, we do not have to take the phenomenological approach adopting 
the exponential type of the heating law. 
Conversely, we can test the validity of the assumption of 
the constant dissipation length for our processes. 

We further explore possibilities whether our model can account for the 
difference between the high-speed solar wind ($\gtrsim 700$km/s) 
from the polar coronal holes and the low-speed solar wind ($\lesssim 400$km/s) 
in the equatorial region 
during the low solar activity phase. We adopt various field strength at the 
bottom of the TR and non-radial expansion factor as well as $\Fwppi$ and 
$\Fwpri$ to study which parameter plays a role in exhibiting the distinctive 
difference between the two types of the solar wind.

\section{Variation of Shock Amplitude}

\subsection{N-waves}
Acoustic waves have not been considered as a major heating source of the 
solar corona for several decades, because those generated by the granule 
motions at photospheric level cannot reach the corona \citep{ss72}. 
However, recent observations have revealed dynamical natures of the corona; 
a lot of transient events such as microflares constantly occur. 
Some models \citep{str99,ks99} describing those events predict excitation 
of acoustic waves, or equivalently slow-mode MHD waves, not at the photosphere 
but in the corona. Moreover, Ofman et al.(1999) identified slow MHD waves 
propagating in polar plumes. 
If one considers such waves travelling along with the magnetic 
field line in the open flux tube region, they eventually form shocks 
and propagate upwardly as the N-waves (Fig.\ref{fig:wvst}), 
a special case of the slow-shock trains, with heating the surrounding 
plasma. For example, the waves of sinusoidal shape with period, 
$\tau_{\parallel}\sim 100$s, steepen their wave fronts to form the shocks 
after propagation by a distance of $\sim 10^4$km in the corona, provided that 
the initial wave amplitude is $\gtrsim 10$\% of the sound speed (S02). 

A model for the N-waves 
was developped for propagation in the chromosphere
by \citet{ss72}, and generalized for propagation in the corona by S02.
Now, we briefly summarize the derivation of an equation for variation of 
amplitude of the N-waves by following S02.
We assume the shock amplitude, $\delta v_{\rm sh, \parallel}$, is sufficiently 
smaller than ambient sound speed, $c_{\rm s}$ (weak-shock approximation). 
Then, entropy generation 
at each shock front is given by the jump condition across the shock as    
\begin{equation}
\label{eq:rhel9}
\frac{\Delta s_{\parallel}}{R}=\frac{\gamma  +1}{12 \gamma ^2} 
(\frac{\Delta p}{p})^3 
=\frac{2\gamma (\gamma +1)}{3} \alpr^3, 
\end{equation}
\citep{LL59}, where $R$ is the gas constant, $\gamma$ is a ratio of the 
specific heat, 
$\Delta p$ denotes pressure difference at the front, 
$p$ is pressure in the upstream region, and 
$\alpr(\equiv \delta_{\rm sh,\parallel}/c_{\rm s})$ is the normalized shock 
amplitude. 
Using the perfect gas law, $T=p/(\rho R)$, the energy loss rate 
$-Q_{\parallel}$ per volume (erg cm$^{-3}$s$^{-1}$) at the 
shocks for the waves with period $\tau_{\parallel}$ is derived from eq. 
(\ref{eq:rhel9}): 
\begin{equation}
\label{eq:hdrs1}
-Q_{\parallel} = - \rho_0 T \Delta s_{\parallel} / \tau_{\parallel} 
= -\frac{2 \gamma (\gamma +1) p_0 
 \alpr^3}{3 \tau_{\parallel}}. 
\end{equation}
Divergence of wave energy, $E_{\lambda_\parallel}=\frac{1}{\varsigma_
{\parallel}}\gamma p \alpr^3 \lambda_{\parallel}$, per wavelength, 
$\lambda_{\parallel}$, for a shape constant, $\varsigma_{\parallel}$($=3$ for 
the N-waves illustrated in Fig.{\ref{fig:wvst}), can be written as 
\begin{equation}
\label{eq:divel}
\mbf{\nabla \cdot E_{\lambda_{\parallel}}}=\frac{dE_{\lambda_{\parallel}}}{dr}
+ \frac{E_{\lambda_{\parallel}}}{A}\frac{dA}{dr}=-Q_{\parallel} 
\tau_{\parallel},
\end{equation}
where $A$ is a cross section of a flow tube.
Substitution of  eq.(\ref{eq:divel}) into the logarithmic derivative of 
$E_{\lambda_{\parallel}}$ gives variation of the shock amplitude in the 
static media as follows (S02) :
\begin{equation}
\label{eq:wvdrrd}
\frac{d\alpr}{dr}=\frac{\alpr}{2}(-\frac{1}{\rho}
\frac{d\rho}{dr}-\frac{\varsigma_{\parallel}}
{3}\frac{2(\gamma +1) \alpr}{c_{\rm s}\tau_{\parallel}} 
-\frac{1}{A}\frac{dA}{dr} -\frac{3}{c_{\rm s}}\frac{dc_{\rm s}}{dr}).
\end{equation}
This equation determines the variation of the shock amplitude of the N-waves 
in the solar corona according to the physical properties of the background 
coronal plasma. 
The first term in the right-hand side denotes the amplification in the 
stratified atmosphere, 
and the second term indicates the loss by the shock heating. 
In the $(1-2)R_{\odot}$ region, where the dissipation is important to  
heat the corona, these two terms always dominate the other terms, 
the third term arising from the geometrical expansion and the fourth from 
temperature variation.

\begin{figure}
\epsfxsize=6cm
\epsfysize=6cm
\epsfbox{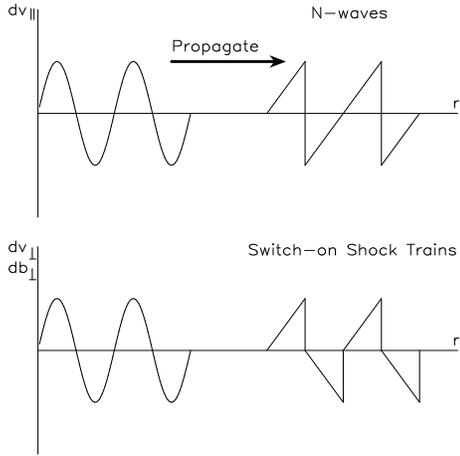}
\caption{Schematic picture of the steepening of acoustic 
waves (top) and linearly polarized \Alfven waves (bottom). 
}
\label{fig:wvst}
\end{figure}

\subsection{Switch-on Shock Trains}
\label{sc:fssh}
The various dynamical activities on the solar surface are expected to 
continuously produce transverse waves. Some of them would be linearly 
polarized \Alfven waves propagating upwardly. Phase speed of the 
linearly polarized \Alfven waves varies as a function of fluctuating field, 
$\delta B_{\perp}$, normal to the background field, $B_{\parallel}$\citep{kl65}:
\begin{equation}
\label{eq:lpalfv}
v_{\rm p}=\frac{\sqrt{B_{\parallel}^2+\delta B_{\perp}^2}}{\sqrt{4\pi\rho}}.
\end{equation} 
The above equation also satisfies the requirement of the plasma 
compressibility \citep{mon59}, 
\begin{equation}
\label{eq:plsmcmp}
\frac{(B_{\parallel}^2+\delta B_{\perp}^2)^{1/2}}{\rho} \simeq {\rm const.}
\end{equation}
Equations (\ref{eq:lpalfv}) and (\ref{eq:plsmcmp}) shows that 
both the wave crest and the wave trough overtake the wave front with a speed,
\begin{equation}
\label{eq:alfpsdf}
\frac{\sqrt{B_{\parallel}^2+\delta B_{\perp}^2}}{\sqrt{4\pi\rho}} - 
\frac{B_{\parallel}}{\sqrt{4\pi\rho_0}} =  
v_{\rm A}[(\frac{B_{\parallel}^2+\delta B_{\perp}^2}
{B_{\parallel}^2})^{\frac{1}{4}} - 1] ,
\end{equation}
where we here define \Alfven velocity, $v_{\rm A}\equiv B_{\parallel}
/\sqrt{4 \pi \rho_0}$ for the ambient density $\rho_0$ at the wave 
fronts. 
Then two shock fronts are formed per wavelength as illustrated 
in Fig. \ref{fig:wvst}, which is also seen in results of time-dependent 
calculations by Nakariakov et al.(2000).   
The figure also displays that as moving from the upstream 
region to the downstream region, $\delta B_{\perp}$ and 
$\delta v_{\perp}$ 'switch-on' from zero to finite values. Therefore, 
it is called switch-on shock, a special case of the fast shocks. Periodic 
trains of the switch-on shocks 
propagate upwardly with a speed $\simeq v_{\rm A}$, similarly to the usual 
\Alfven waves except that they dissipate the wave energy at each shock.

Let us estimate a distance the sinusoidal wave travels till the formation 
of the shock fronts. 
We can categorize 
these transverse waves into two types in terms of location of the wave 
generation.
First type is the wave generated by the granule motions 
at the photosphere. 
Since the \Alfven velocity in the photosphere and low chromosphere is 
quite small, the sinusoidal waves steepen to form the shock trains 
before reaching the corona. 
This is seen in recent time-dependent 
calculation by \citet{ks99}, illustrating that the transverse waves created 
at the photosphere propagate into the corona as the fast shocks. 

Second type is the wave excited by transient events such as microflares in 
the corona.  To form the front, the wave crest or trough have to 
overtake the wave front by a quarter of a wave length, $\lambda_{\perp}$. 
If we set that the sinusoidal waves are generated at time $t_0$ 
and position $R_0$,  
time $t$ and position $R$, at which the waves steepen to form the fronts, 
satisfy the conditions below in a static atmosphere. 
\begin{equation}
\label{eq:dsalfst}
\hspace{-5mm}\int_{t_0}^t dt' v_{\rm A}
[(\frac{B_{\parallel}^2+\delta B_{\perp}^2}
{B_{\parallel}^2})^{\frac{1}{4}} - 1] =
 \int_{R_0}^R dr [(\frac{B_{\parallel}^2+\delta B_{\perp}^2}
{B_{\parallel}^2})^{\frac{1}{4}} - 1] =\frac{\lambda_{\perp}}{4}
\end{equation}
If we consider wave with period, $\tau_{\perp}=120$s, generated at 
$R_0=R_{\odot} + 1\times 10^4$km, substitution of typical values in the 
corona of 
$B_{\parallel}\sim 5$G, $\delta B_{\perp}\sim 1$G, and density 
scale height, $H\simeq 6\times 10^4(\frac{T}
{10^6 {\rm K}})$km,  into eq.(\ref{eq:dsalfst}) gives 
$R\simeq 1.3R_{\odot}$. 
It indicates that even if the linearly polarized \Alfven waves 
are excited in the corona, they form the fast shocks in the low corona. 
Moreover, the assumption of the generation of the sinusoidal waves is 
idealistic and the initial wave shape is supposed to be more or less out of 
shape in the real solar corona. In a such case the distance till 
the formation of the shocks should become shorter.      

From now, we would like to derive an equation describing variation of 
amplitude of the switch-on shock trains by extending the formulation by H82. 
From the jump conditions across the switch-on shock front \citep{bs69}, 
entropy generation $\Delta s_{\perp}$ at the shock can be expressed as 
\begin{equation}
\label{eq:jpfs1}
\frac{\Delta s_{\perp}}{R}\simeq \frac{B_{\parallel}^2}{8 \pi p}(\sigma -1)^2
\end{equation}
for the weak-shock condition of $\sigma -1 \ll 1$ (H82), 
where $p$ is pressure in the upstream region and $\sigma$ is a ratio of 
the downstream density to the upstream density. 
Field strength, $\delta B_{\perp}$, in the downstream region perpendicular 
to the underlying field, $B_{\rm \parallel}$, 
can also be derived from the shock conditions: 
\begin{equation}
\label{eq:jpfs2}
\delta B_{\perp}^2 \simeq 2 (\sigma -1)B_{\parallel}^2 
(1-c_{\rm s}^2/v_{\rm A}^2),
\end{equation}
where $c_{\rm s}$ and $v_{\rm A}$ is the sound speed and \Alfven speed in the 
upstream region.
Eliminating $(\sigma -1)$ from eqs.(\ref{eq:jpfs1}) and (\ref{eq:jpfs2}), 
$\Delta s_{\perp}$ is rewritten as a function of $\delta B_{\perp}$ or 
$\delta v_{\perp}$:
\begin{eqnarray}
\label{eq:etrpsw2}
\frac{\Delta s_{\perp}}{R} &\simeq & \frac{B_{\parallel}^2}{32\pi p 
(1-c_{\rm s}^2/v_{\rm A}^2)^2}(\frac{\delta B_{\perp}}
{B_{\parallel}})^4 \\
& = & \frac{B_{\parallel}^2}{32\pi p_1 
(1-c_{\rm s}^2/v_{\rm A}^2)^2}(\frac{\delta v_{\perp}}{v_{A}})^4 \\
\label{eq:etrpsw3}
& \equiv &  \frac{B_{\parallel,1}^2}{32\pi p_1 
(1-c_{\rm s,1}^2/v_{\rm A,1}^2)^2}(\alpp)^4,
\end{eqnarray}
where we have used the relation of $\delta v_{\perp}=\delta b_{\perp}/
\sqrt{4\pi \rho}$ in the downstream region.
Recalling that the two shock fronts form per wavelength in this case (H82; 
Fig.\ref{fig:wvst}), 
energy loss rate $-Q_{\perp}$(erg cm$^{-3}$s$^{-1}$) 
for the waves with period $\tau_{\perp}$ becomes 
\begin{equation}
\label{eq:htswsh}
-Q_{\perp}= -\rho T \Delta s_{\perp} /(\frac{1}{2}\tau_{\perp}) = 
-\frac{B_{\parallel,1}^2}{16 \pi \tau_{\perp}} \frac{\alpp^4}{(1-c_{\rm s}^2/v_{\rm A}^2)^2} , 
\end{equation}
It should be noted that the heating, $Q_{\perp}$, is proportional to 
$\alpp^4$, which is a higher order term with respect to $\alpha_{\rm sh}$ 
compared to the case of the N-waves giving 
$Q_{\parallel} \propto \alpr^3$ 
(eq.(\ref{eq:hdrs1})), since the liberated energy at the switch-on shocks 
is transfered to magnetic field as well as heat. 
This fact implies that the damping by the switch-on 
shocks is weaker and the periodic trains could travel to a more distant region 
than the N-waves to contribute to the heating in the outer region. 

Now we derive the equation for $\alpp$ by using 
eq.(\ref{eq:htswsh}) under the WKB approximation.
In moving media with velocity $v$, even though the dissipation does not 
occur, the wave energy flux is not conserved. Instead, one can have a new 
conserved quantity of wave action constant which is defined as \citep{jaq77}
\begin{equation}
\label{eq.aevacct}
\mbf{\Spp} \equiv \frac{\rho  \alpp^2 
v_{\rm A}(v + v_{\rm A})(\mbf{v + v_{\rm A}})}{\varsigma_{\perp} },
\end{equation}
where $\varsigma_{\perp}$ is a shape constant for the switch-on shock trains. 
In our calculations we set $\varsigma_{\perp}=3$, corresponding to the wave 
illustrated in Fig.\ref{fig:wvst}.
Although $\mbf{\Spp}$ is the conserved quantity without dissipation, 
we have to 
take into account the energy loss at the shocks for the switch-on shock 
trains. Then, an equation governing variation of $\mbf{\Spp}$ becomes 
\begin{equation}
\label{eq:alfvht1}
\frac{v_{\rm A}}{v_{\rm A}+v} \mbf{\nabla \cdot \Spp}  
= -Q_{\perp}=-\frac{B_{\parallel}^2}{16 \pi \tau_{\perp}} \frac{\alpp^4}
{(1-c_{\rm s}^2/v_{\rm A}^2)^2} .
\end{equation}
Since the left-hand side of eq.(\ref{eq:alfvht1}) is rewritten in terms of 
wave energy flux, $\mbf{\Fwpp}$, and wave pressure, $p_{\rm w,\perp}$, as 
$\frac{v_{\rm A}}{v_{\rm A}+v}\mbf{\nabla \cdot \Spp} = 
\mbf{\nabla\cdot\Fwpp}-\mbf{v\cdot \nabla}p_{\rm w,\perp} $, 
eq.(\ref{eq:alfvht1}) essentially indicates variation of the wave energy flux. 
In eq.(\ref{eq:alfvht1}) wave period, $\tau_{\perp}$, is measured 
from the observer moving with the flow. 
It is more useful to consider the period, 
\begin{equation}
\tau_{\rm \perp,i} = \tau_{\perp} v_{\rm A}/(v_{\rm A}+v),  
\end{equation}
as seen by the stationary inertial observer, since it is a constant 
along with a given field line (H82).    
Therefore, eq.(\ref{eq:alfvht1}) is rewritten as 
\begin{equation}
\label{eq:alfvht2}
\frac{1}{A}\frac{d}{dr}(A \Spp) 
= -\frac{B_{\parallel}^2}{16 \pi \tau_{\rm \perp,i}} 
\frac{\alpp^4}{(1-c_{\rm s}^2/v_{\rm A}^2)^2}, 
\end{equation}
where we here assume $\mbf{\Spp}$ has the only radial variation:
$\mbf{\nabla \cdot \Spp}=\frac{1}{A}\frac{d(A\Spp)}{dr}$;
$\Spp\equiv|\mbf{\Spp}|$.
The logarithmic derivative of $\Spp$ (eq.(\ref{eq.aevacct})) gives 
\begin{displaymath}
\frac{1}{\Spp}\frac{d\Spp}{dr}=\frac{1}{\rho}\frac{d\rho}{dr}+
\frac{2}{\alpp}\frac{d\alpp}{dr}
\end{displaymath}
\begin{equation}
\label{eq:lgdalf}
\hspace{2cm}
+\frac{3v_{\rm A} + v}{v_{\rm A}(v_{\rm A} + v)}\frac{d v_{\rm A}}{dr}
+\frac{2}{v_{\rm A} + v}\frac{d v}{dr}.
\end{equation}
Then, an equation describing variation of the shock amplitude in the steady 
flow is derived from eqs. (\ref{eq:alfvht2}) and (\ref{eq:lgdalf}):
\begin{displaymath}
\frac{d\alpp}{dr}=\frac{\alpp}{2} 
[-\frac{1}{\rho}\frac{d\rho}{dr}- \frac{\varsigma_{\perp} B_{\parallel} 
\alpp^2}
{8 \tau_{\rm \perp,i}\sqrt{\pi \rho}(v_{\rm A}+v)^2
(1-c_{\rm s}^2/v_{\rm A}^2)^2}
\end{displaymath}
\begin{equation}
\label{eq:alfwvamp1}
-\frac{1}{A}\frac{dA}{dr}-\frac{3v_{\rm A} + v}{v_{\rm A}(v_{\rm A} + v)}
\frac{d v_{\rm A}}{dr} -\frac{2}{v_{\rm A} + v}\frac{dv}{dr}].
\end{equation}
We would like to emphasize that eq.(\ref{eq:alfwvamp1}) is derived 
for the first time, whereas eq.(\ref{eq:alfvht2}) is directly solved 
in the fixed background structures in H82. 
An advantage of using eq.(\ref{eq:alfwvamp1}) instead of eq.(\ref{eq:alfvht2}) 
is that we can directly get the shock strength, $\alpp$, 
when performing the numerical integration.

In the static limit ($v \ll v_{\rm A}$), eq.(\ref{eq:alfwvamp1}) shows  
an analogous form to eq.(\ref{eq:wvdrrd}) for the N-waves: 
\begin{equation}
\label{eq:alfwvamp2}
\hspace{-5mm}\frac{d\alpp}{dr}=\frac{\alpp}{2} 
[-\frac{1}{\rho}\frac{d\rho}{dr} - \frac{\varsigma_{\perp}  \alpp^2}
{4 \tau_{\perp}  v_{\rm A}(1-c_{\rm s}^2/v_{\rm A}^2)^2} 
-\frac{1}{A}\frac{dA}{dr}-\frac{3}{v_{\rm A}}\frac{d v_{\rm A}}{dr}]. 
\end{equation}
In the square brackets of the right-hand side, the 1st and 3rd terms are 
identical 
to those in eq.(\ref{eq:wvdrrd}), and in the 4th term $v_{\rm A}$ 
appears instead of $c_{\rm s}$.  An important difference is seen in the 2nd 
term denoting the dissipation at the shocks. The variation of the amplitude is 
proportional to $\alpp^3$ in the case of the switch-on shocks, while that to 
$\alpr^2$ in the N-wave case, which reflects the different dependencies of 
the entropy generation on $\alpha_{\rm sh}$ (eqs.(\ref{eq:rhel9}) \& 
(\ref{eq:etrpsw3})). 
This also shows that the switch-on shock trains are less dissipative.

\subsection{Limitations in Our Modelling}
We had better comment on the limitation when using eqs.(\ref{eq:wvdrrd}) 
and (\ref{eq:alfwvamp1}) for the analysis of the propagation of the shock 
trains. 
An interaction between the N-waves and the switch-on shock trains is 
supposed to give little influence, since it is purely the non-linear 
effect. Our calculations show that the shock strength is finite 
but still small ($\alpha_{\rm sh}< 1$) (\S\ref{sc:obs}). 
Consequently, the two modes of the shock trains can travel almost 
independently, hence, it is justified to solve eqs.(\ref{eq:wvdrrd}) 
and (\ref{eq:alfwvamp1}) separately.   

The first limitation on the propagation of the N-waves is that it is valid 
only in the static media. However, this is accomplished in our calculation 
since the N-waves 
with $\tau_{\parallel}\sim 100$s mostly dissipate in the low corona where 
the static approximation is sufficient (\S\ref{sc:obs}). 
The second limitation is that it does not take into account the effect of 
the gravity. This is 
also negligible for the waves with $\tau_{\parallel}\sim 100$s, because 
the acoustic cut-off period for the acoustic gravity wave is much greater 
$(\tau_{\rm ac}\gtrsim 1000{\rm s})$ in the corona.      
The third limitation is that $\alpr$ should be less than 1, since 
eq.(\ref{eq:wvdrrd}) is derived under the weak shock approximation. 
As shown in the results, our model does not break this limitation either, 
whereas the condition at the coronal base is almost marginal, 
$\alpr\simeq 0.8$ (Fig.\ref{fig:twlw}). 
The final limitation is that we do not take into account viscosity explicitly. 
Although we consider viscosity implicitly through the shock, 
%(shock fronts form as a result of viscosity), 
compressive viscosity, which is not 
negligible in the corona, may make the acoustic waves dissipate even when 
they mildly steepen before the shock formation (Ofman et al.2000). 
However, this effect dose not affect the coronal 
structure, since the dissipated energy from the acoustic waves is mainly lost 
as the downward thermal conduction, unrelated to how rapidly they dissipate 
(see \S\ref{sc:dpip}). 

As for the switch-on shock trains, we do not have to consider the problem 
in the static media as eq.(\ref{eq:alfwvamp1}) has been derived with taking 
into account the steady flow.
The effect of the gravity is negligible, since the switch-on shock trains are 
basically transverse. The condition of the weak shock, $\alpp <1$, is 
also accomplished (Fig.\ref{fig:twlw}). 
The necessary condition for the WKB approximation is that the scale height 
of the \Alfven speed, $H_{\rm A}$, is sufficiently larger than the wavelength. 
The condition is easiest to be violated in the low corona 
where $H_{\rm A}\sim 2\times 10^5$km and 
$v_{\rm A} \sim 10^3$km/s. Therefore, the WKB approximation becomes marginal 
there for the waves with $\tau_{\perp}\simeq 200$s, while it is 
reasonable for those with $\tau_{\perp}\lesssim 100$s.   
In more sophisticated models dynamical treatments (e.g. Ofman \& Davila 1997; 
1998) are required for those low-frequency waves. 
The most severe limitation is that we assume that the transverse waves 
dissipate only through the switch-on shocks. 
Although taking into account multi-dimensional effects is beyond the scope of 
this paper, transverse variation of the field strength leads to 
the dissipation by the phase mixing 
\citep{hp83,sg84} and through the mode conversion \citep{nrm97}.
If the waves are not coherent and show more or less 
turbulent-like natures, the turbulent cascade might occur (e.g. Hollweg 1986; 
Hu, Habbal, \& Li 1999).
To determine the dominant process(es) in the solar corona and the solar wind 
is an important task. By comparing our results 
with those considering the other processes, 
we can proceed a further step 
to understand the coronal heating and the solar wind acceleration.

\section{Model for Corona and Solar Wind}
\subsection{Basic Equations}
In this study, the solar corona and wind are assumed to be fully ionized MHD 
plasma consisting of protons and electrons. The fluid (MHD) treatment might be 
controversial especially in the outer region ($r\gtrsim 10 R_{\odot}$) as 
the plasma becomes collisionless with respect to the Coulomb collision, and 
the kinetic treatment would be required in more realistic models.  
However, 
randomness of the magnetic fields may operate to let them exhibit the MHD 
characteristics.

We here present basic equations to describe 
coronal wind structure in a flow tube with a cross-section of $A$ 
under a steady state condition. 
The equation of mass conservation is written as 
\begin{equation}
\label{eq:cntn}
\rho v A = {\rm const.},
\end{equation}
where mass density, $\rho$, is related to proton and electron number density, 
$n_{\rm p}$ and $n_{\rm e}$, and proton and electron mass, $m_{\rm p}$ and 
$m_{\rm e}$, as $\rho=n_{\rm p} m_{\rm p}+n_{\rm e} m_{\rm e}\simeq 
n_{\rm p} m_{\rm p}$.
We assume $n_{\rm e}=n_{\rm p}\equiv n$, and the proton 
and electron components have the equal outflow velocity in the solar wind. 
Then, the equation of momentum conservation becomes
\begin{equation}
\label{eq:eqm}
v \frac{dv}{dr}=-\frac{G M_{\odot}}{r^2}-\frac{1}{\rho}\frac{dp}{dr} 
-\frac{1}{\rho}\frac{d p_{\rm w,\perp}}{dr}
-\frac{1}{\rho}\frac{d p_{\rm w,\parallel}}{dr},
\end{equation}
where $p_{\rm w,\perp}$ is the wave pressure of the switch-on shock trains and 
$p_{\rm w,\parallel}$ is that of the N-waves.
Gas pressure, $p$, is related to $\rho$ and  
temperature of the protons, $T_{\rm p}$, and the electrons, $T_{\rm e}$ 
by an equation of state for an ideal gas:
\begin{equation}
\label{eq:eqst}
p=\rho \frac{k_{\rm B}}{m_{\rm p}}(T_{\rm p}+T_{\rm e}) ,
\end{equation}
where $k_{\rm B}$ is the Boltzmann constant.

As for the energy transfer, the protons and electrons are allowed to have 
different temperature. 
The energy equation for the protons is 
\begin{equation}
\label{eq:engprt}
\frac{n^{\gamma}}{\gamma -1} v \frac{d}{dr}
(\frac{k_{\rm B} T_{\rm p}}{n^{\gamma -1}}) = 
-\mbf{\nabla\cdot F_{\rm c,p}}+C_{\rm pe} + Q_{\perp,{\rm p}}
+Q_{\parallel,{\rm p}},
\end{equation}
and that for the electrons is 
\begin{equation}
\label{eq:engelc}
\frac{n^{\gamma}}{\gamma -1} v \frac{d}{dr}
(\frac{k_{\rm B} T_{\rm e}}{n^{\gamma -1}}) = 
-\mbf{\nabla\cdot F_{\rm c,e}}+C_{\rm ep} + Q_{\perp,{\rm e}}+
Q_{\parallel,{\rm e}} - n^2\Phi(T_{\rm e}),
\end{equation}
where $C_{\rm pe} = -C_{\rm ep}$ 
indicate energy exchange between the electrons and protons by Coulomb 
collision\citep{brg65,sl94}, $Q_{\perp,{\rm p}}$ and $Q_{\perp,{\rm e}}$ 
denote heating from the switch-on shock trains to the protons and electrons, 
$Q_{\parallel,{\rm p}}$ and $Q_{\parallel,{\rm e}}$ are heating from the 
N-waves to the protons and 
electrons, and $\Phi(T_{\rm e})$ is radiative loss function by \citet{LM90}.
$\mbf{F_{\rm c,p}}$ and $\mbf{F_{\rm c,e}}$ is thermal conduction taken by 
the protons and electrons, whereas we here use the classical Spitzer 
conductivity by the Coulomb collisions:
\begin{equation}
F_{\rm c,p}=-\kappa_{\rm p} \frac{dT_{\rm p}}{dr} = -\kappa_{\rm p,0} 
T_{\rm p}^{5\over{2}}\frac{dT_{\rm p}}{dr},
\label{eq.cndp}
\end{equation}

\begin{equation}
F_{\rm c,e}=-\kappa_{\rm e} \frac{dT_{\rm e}}{dr} 
= -\kappa_{e,0} T_{\rm e}^{5\over{2}}\frac{dT_{\rm e}}{dr},
\label{eq.cnde}
\end{equation}
where $\kappa_{p,0}=3.2\times 10^{-8}$ 
and $\kappa_{e,0}=7.8\times 10^{-7}$(erg cm$^{-1}$s$^{-1}$K$^{-7/2}$) 
\citep{brg65}.
It should be noted that adoption of the above classical forms of thermal 
conduction is controversial especially in the outer corona and the solar 
wind region, since the plasma becomes collisionless there (e.g. Hollweg 1976).

In the very inner corona where the collisional coupling between the protons 
and electrons is satisfied, we use one-fluid approximation for the 
energy transfer. The energy equation for the one-component plasma can be 
obtained by combining eqs.(\ref{eq:engprt}) and (\ref{eq:engelc}): 
\begin{equation}
\label{eq:engtt}
\frac{2n^{\gamma}}{\gamma -1} v \frac{d}{dr}
(\frac{k_{\rm B} T}{n^{\gamma -1}}) = 
-\mbf{\nabla\cdot F_{\rm c}} + Q_{\perp}+Q_{\parallel} - n^2\Phi(T_{\rm e}),
\end{equation}
where $T\equiv(T_{\rm p}+T_{\rm e})/2$, $F_{\rm c}\simeq -(\kappa_{\rm p,0}+
\kappa_{\rm e,0}) 
T^{5\over{2}}\frac{dT}{dr}$, $Q_{\perp}=Q_{\rm \perp,p}+Q_{\rm \perp,e}$, and 
$Q_{\parallel}=Q_{\rm \parallel,p}+Q_{\rm \parallel,e}$. Please note that 
total number density becomes $2n(=n_{\rm p}+n_{\rm e})$. 
A practical way to connect the coronal wind structures of the two-component 
plasma to those of the one-component plasma will be described in next 
subsection.     

For the wave pressure terms in eq.(\ref{eq:eqm}), 
we take the WKB approximation.
Wave pressure of the switch-on shock trains is essentially 
identical to that of the usual \Alfven waves. Then, the term as to 
the wave pressure gradient in eq.(\ref{eq:eqm}) can be 
written as  
\begin{equation}
\frac{dp_{\rm w,\perp}}{dr}=\frac{d}{dr}(\frac{\delta B_{\perp}^2}
{8\pi\varsigma_{\perp}})
=\frac{d}{dr}(\frac{\alpp^2 B_{\parallel}^2}{8\pi\varsigma_{\perp}} )
\label{eq.wvprss}
\end{equation} 
\citep{jaq77}.
Although the wave pressure term for the N-waves can be found in 
S02, we neglect it since inclusion of this term does not affect the results 
at all (S02):
\begin{equation}
\frac{dp_{\rm w,\parallel}}{dr}=0 .
\end{equation} 
The heating terms, $Q_{\perp}$ and $Q_{\parallel}$, have been already 
derived in eqs.(\ref{eq:alfvht2}) and (\ref{eq:hdrs1}).
Distribution of the dissipated energy at the shocks to the protons and 
electrons is unclear. Therefore, we use following parameterization : 
\begin{equation}
Q_{\perp/\parallel,{\rm p}}=(1-h_{\perp/\parallel}) Q_{\perp/\parallel}
\end{equation}
\begin{equation} 
Q_{\perp/\parallel,{\rm e}}=h_{\perp/\parallel} Q_{\perp/\parallel} .
\end{equation}
We perform our calculation for two extreme cases of $h_{\perp/\parallel} = 
0.5$ and 0, whereas we mainly discuss model results adopting 
$h_{\perp/\parallel}=0.5$.

Since the switch-on shock trains travel in the very similar manner to 
the usual \Alfven waves except the shock dissipation, 
wave energy flux, $\mbf{\Fwpp}$, is explicitly expressed, 
following \citet{jaq77}, as 
\begin{eqnarray}
\label{eq:trwven}
\mbf{\Fwpp}&=&\frac{1}{\varsigma_{\perp}}\rho \alpp^2 v_{\rm A}^2 [
\mbf{v}_{\rm A}(1+\alpp^2)^{\frac{1}{4}}+\frac{3}{2}\mbf{v}] \nonumber \\
&\simeq& \frac{1}{\varsigma_{\perp}}\rho \alpp^2 v_{\rm A}^2 [
\mbf{v}_{\rm A}+\frac{3}{2}\mbf{v}].
\end{eqnarray} 
For the N-waves, we use a form in the static media (S02),   
\begin{equation}
\label{eq:lgwven}
\mbf{\Fwpr}\simeq\frac{1}{\varsigma_{\parallel}}\rho \alpr^2 c_{\rm s}^2 
\mbf{c}_{\rm s}(1+\frac{\gamma +1}{2}\alpr),
\end{equation} 
since they dissipate quickly (\S\ref{sc:prwv}) and 
work effectively only in the low corona where the static approximation is 
fulfilled.  

To take into account the non-radial 
expansion of the flow tube due to configurations of the magnetic field, 
the cross-sectional area, $A$, is modeled as 
\begin{equation}
\label{eq:ftg}
A=r^2 \frac{f_{\rm max}e^{(r-r_1)/\sigma}+f_1}{e^{(r-r_1)/\sigma}+1}
\end{equation}
where
$$
f_1=1-(f_{\rm max}-1)e^{(1-r_1)/\sigma}
$$
\citep{ko76,wtb88}. 
The cross section expands from unity to $f_{\rm max}$ most drastically between 
$r=r_1-\sigma$ and $r_1+\sigma$. 
Of the three input parameters, $f_{\rm max}$ 
is the most important in determining the solar wind structure. In this paper, 
we consider cases between $f_{\rm max}=1$ and $f_{\rm max}=20$. 
As for the other two 
parameters, we employ $r_1=1.25R_{\odot}$ and $\sigma=0.1R_{\odot}$ 
\citep{wtb88}.

The geometrical expansion of the flow tubes is subject to the open magnetic 
field lines. Conservation of magnetic fields, $\mbf{\nabla \cdot B}=0$, gives 
a condition for radial magnetic field, $B_{\parallel}$, as
\begin{equation} 
\label{eq:divb}
B_{\parallel}A={\rm const.}
\end{equation}
In our modelling, we leave the radial field strength, $B_0$, at the 
inner boundary located at the bottom of the TR as an input 
parameter, and determine $B_{\parallel}(r)$ according to eq.(\ref{eq:divb}).
Note that $B_0$ is supposed to be smaller than the field strength at the 
photosphere, since the field lines are supposed to open out with height even 
in the chromosphere (e.g. Giovanelli 1980). 

\subsection{Boundary Conditions and Numerical Method}
Now we would like to explain the practical aspects of our method of 
constructing a unique coronal 
wind structure with respect to various input properties of the linearly 
polarized \Alfven waves and the acoustic waves.
In order to solve both the heating of the corona (energy transfer) and the 
acceleration of the solar wind (momentum transfer) consistently, 
our calculation is performed 
in a broad region from the bottom of the TR 
where temperature, $T_{\rm p}=T_{\rm e}=T_{\rm in}=2\times 
10^4$K, located at $r_{\rm in}=1.003R_{\odot}$ (e.g. Golub \& Pasachoff 1997) 
to an arbitrary outer boundary at $r_{\rm out}=215 R_{\odot}(=1{\rm AU})$.

Only transonic solutions are allowed for the flow speed, $v$. 
Numerical integration of eqs.(\ref{eq:wvdrrd}),(\ref{eq:alfwvamp1}), 
(\ref{eq:eqm}), (\ref{eq:engprt}), and (\ref{eq:engelc}) is carried out 
simultaneously from the critical point, $r=r_{\rm cr}$, defined below 
by Rosenbrock method, which adopts implicit spatial grids to keep the
 numerical stability \citep{ptvf92}. 
When we perform the integration of the energy equations 
(\ref{eq:engprt}) and (\ref{eq:engelc}), isothermal sound velocities for 
the protons, $a_{\rm p}^2=\frac{k_{\rm B}T_{\rm p}}{m_{\rm p}}$, and for 
the electrons, $a_{\rm e}^2=\frac{k_{\rm B}T_{\rm e}}{m_{\rm e}}$, are 
used. To start the integration, we set seven initial guesses,  
$\rho_{\rm cr}$, $a_{\rm p,cr}^2$, $a_{\rm e,cr}^2$, 
$(da_{\rm p}^2/dr)_{\rm cr}$, $(da_{\rm e}^2/dr)_{\rm cr}$, 
$\alpha_{\rm sh,\perp,cr}$, and 
$\alpha_{\rm sh,\parallel,cr}$ (subscript, 'cr' denotes the 
'critical' point). $r_{\rm cr}$ and $v_{\rm cr}$ are automatically derived 
from conditions that the numerator and denominator of a transformed form of 
the momentum equation (\ref{eq:eqm}),
\begin{displaymath}
\frac{dv}{dr}=[v - \frac{a^2}{v}-\frac{\alpp^2}{4\varsigma_{\perp}}
\frac{v_{\rm A}^2(v_{\rm A}+3v)}{(v_{\rm A}+v)v}]^{-1} 
\end{displaymath}
\begin{displaymath}
[(a^2+\frac{\alpp^2}{4\varsigma_{\perp}}
\frac{v_{\rm A}^2(v_{\rm A}+3v)}{v_{\rm A}+v})
\frac{1}{A}\frac{dA}{dr}-\frac{da^2}{dr}-\frac{GM_{\odot}}{r^2}
\end{displaymath}
\begin{equation}
\label{eq:eqmtr}
+\frac{\alpp^4 v_{\rm A}^3}{8\tau_{\rm \perp,i}(v+v_{\rm A})^2
(1-c_{\rm s}^2/v_{\rm A}^2)^2}] ,
\end{equation}
should become zero at $r=r_{\rm cr}$, where 
$a^2\equiv a_{\rm p}^2+\frac{m_{\rm e}}{m_{\rm p}}a_{\rm e}^2$. 
$(dv/dr)_{\rm cr}$ is also obtained by differentiating both the 
numerator and denominator of eq.(\ref{eq:eqmtr}) (De l'Hopital's rule; see
e.g. Lamers \& Cassinelli 1999). 

The integration is firstly carried out to outward direction to satisfy 
following outer boundary conditions by tuning $(da_{\rm p}^2/dr)_{\rm cr}$ and 
$(da_{\rm e}^2/dr)_{\rm cr}$:
\begin{equation}
\label{eq:bc41}
\mbf{\nabla \cdot F}_{\rm c,p}(r_{\rm out})=0,
\end{equation}
\begin{equation}
\label{eq:bc42}
\mbf{\nabla \cdot F}_{\rm c,e}(r_{\rm out})=0.
\end{equation}
Although the integration is to be carried on to 
$r\rightarrow \infty$ idealistically, 
we have to stop it at $r=r_{\rm out}$.  
The above conditions are requirements to keep further outward integration 
stably without divergent behaviors of $T\rightarrow 0$ or 
$T\rightarrow \infty$.
When performing the outward integration, we have to be careful when solving eq.
(\ref{eq:alfwvamp1}) for the switch-on shocks, since it is only valid 
in the low-$\beta$ plasma ($v_{\rm A}> c_{\rm s}$).
(Note that the second term of the right-hand side 
includes $1/(1-c_{\rm s}^2/v_{\rm A}^2)$.) The condition of 
$v_{\rm A}> c_{\rm s}$ holds in a range of $r\lesssim (40-120)R_{\odot}$ in 
our calculations. 
We firstly solve eq.(\ref{eq:alfwvamp1}), and once the condition of the 
low-$\beta$ breaks, we set $\alpp=0$ instead of solving eq.
(\ref{eq:alfwvamp1}). 
This ad hoc treatment 
can be justified, since the wave energy of the switch-on shocks 
is almost completely dissipated in the low-$\beta$ region.

After fulfilling the outer boundary conditions, 
eqs.(\ref{eq:bc41}) and (\ref{eq:bc42}), 
the inward integration is followed. 
If our initial guesses are good, $T_{\rm p}$ and $T_{\rm e}$ approach 
each other as $r$ decreases. Therefore, to satisfy the condition of 
\begin{equation}
\label{eq:bc5}
T_{\rm p}=T_{\rm e} \; \; {\rm for}\; \;  n > n_{\rm c},
\end{equation}
we look for the correct $a_{\rm p,cr}^2$ and $a_{\rm e,cr}^2$, where 
$n_{\rm c}(\simeq 3\times 10^7{\rm cm}^{-3})$ is coupling density above 
which the thermal coupling between the protons and electrons is well-achieved.
When $n>n_{\rm c}$ is realized, 
the term for the  energy exchange between the protons and 
electrons dominates the other terms in eqs.(\ref{eq:engprt}) and 
(\ref{eq:engelc}).
The solar corona is treated as the one-component plasma in the inner part.
In our calculations, the two-component plasma is converted to the 
one-component plasma around $1.1 - 1.4R_{\odot}$.

Then, we integrate eq.(\ref{eq:engtt}) inwardly instead of 
eqs.(\ref{eq:engprt}) and (\ref{eq:engelc}) to $r=r_{\rm in}$ to satisfy 
the inner boundary conditions below:  
\begin{equation}
\label{eq:bc11}
\Fwpp(r_{\rm in})=\Fwpp(r_{\rm \perp, d})=\Fwppi
\end{equation}
\begin{equation}
\label{eq:bc12}
\Fwpr(r_{\rm in})=\Fwpr(r_{\rm \parallel,d})=\Fwpri
\end{equation}
\begin{equation}
\label{eq:bc2}
T(r_{\rm in})=T_{\rm in},
\end{equation}
\begin{equation}
\label{eq:bc3}
|F_{\rm c}(r_{\rm in})|(\simeq 0) \ll |F_{\rm c, max}|, 
\end{equation}
where $F_{\rm c, max}$ in eq.(\ref{eq:bc3})  
is the maximum value of the 
downward conductive flux in the inner corona.
The first and second conditions denote that the wave energy flux  
of the linearly polarized \Alfven waves (transverse mode) and acoustic waves 
(longitudinal mode) must 
agree with the given values when the waves start to dissipate. 
$r_{\rm \perp, d}$ and $r_{\rm \parallel,d}$ are locations where the waves 
start to dissipate. For the acoustic waves, we set 
$r_{\rm \parallel,d}=R_{\odot}+2\times10^4$km, focusing on the waves with 
$\tau_{\parallel}\sim 100$s excited 
at $R_{\odot} +10^4$km in the corona by some dynamical events (S02). 
For the linearly polarized \Alfven waves, we choose $r_{\rm \perp, d}=
r_{\rm in}$, which corresponds to the case that the waves are generated at 
the photosphere and start to dissipate from the bottom of the corona. 
If one takes the waves excited in 
the corona, larger $r_{\rm \perp, d}$ should be adopted. For example, 
$r_{\rm \perp, d}\simeq 1.3R_{\odot}$ for the sinusoidal waves with 
$\tau_{\perp}=120$s (\S\ref{sc:fssh}). However, the 
dissipation of the switch-on shock trains in the low corona 
($r\lesssim 1.5R_{\odot}$) is not significant 
(Fig.\ref{fig:twlw}), hence, the choice of 
different $r_{\rm \perp, d}$s affects the results little. 

The third condition, eq.(\ref{eq:bc2}), is straightforward; the 
temperatures have to coincide with the 
fixed value at the inner boundary. 
The fourth condition, eq.(\ref{eq:bc3}), is the requirement that the 
downward thermal 
conductive flux should radiate away and become sufficiently small at the 
bottom of the TR ($T=2 \times 10^4$K), diminishing from its enormous value 
at the coronal base ($T\sim 10^6$K). 
Practically, we continue calculations iteratively until 
$F_{\rm c}(r_{\rm in})/F_{\rm c, max}<1\%$ is satisfied.
Note that thanks to this condition, the coronal base 
density, which is poorly determined from the observations, does not have to 
be used as a boundary condition \citep{hm82a,hm82b,wtb88}. 
The density at the coronal base or the TR is calculated as an output; 
a larger input energy increases downward conductive flux from the lower 
corona to the chromosphere, demanding a larger density in the 
coronal base and TR to enhance radiative cooling to balance 
with the increased conductive heating. 

Unless the above boundary conditions of eqs.(\ref{eq:bc11}) -- 
(\ref{eq:bc3}) are simultaneously satisfied, one returns to the 
outward integration with preparing an improved set of initial 
guesses and iteratively determine the unique coronal wind structure 
on the given wave energy flux. Physically, the 
conditions of the wave flux (eqs. (\ref{eq:bc11}) and (\ref{eq:bc12})) 
regulates $\alpha_{\rm sh,\perp,cr}$ and $\alpha_{\rm sh,\parallel,cr}$, 
those of the temperature (eqs.(\ref{eq:bc5}) and (\ref{eq:bc2})) 
regulates $a_{\rm p,cr}^2$ and $a_{\rm e,cr}^2$, and that of 
the conductive flux (eq.(\ref{eq:bc3})) regulates 
$\rho_{\rm cr}$. These relations guide the improvement of the respective 
initial guesses, though they are not independently approved. 
Finally, we can determine the unique coronal wind structure by iteratively 
improving the seven initial guesses to fulfill the seven conditions of 
eqs.(\ref{eq:bc41})-(\ref{eq:bc3}).

\section{Dependency on Input Parameters}
\label{sc:dpip}
We inspect dependency of the results 
on the input parameters, $\Fwppi$, $\Fwpri$, $\tau_{\perp}$, 
$\tau_{\parallel}$, $f_{\rm max}$ and $B_0$ to prepare 
for the next section in which we explore the model parameters explaining 
the observed high- and low-speed solar wind. 
From now, we express period of the switch-on shock trains seen by the 
stationary observer just as $\tau_{\perp}$ instead of $\tau_{\rm \perp,i}$. 
We show dependencies of six characteristic 
properties, proton flux, $(n_{\rm p}v)_{\rm 1AU}$, at 1AU, solar wind speed, 
$v_{\rm 1AU}$, at 1AU, peak temperature of the protons, $T_{\rm p,max}$, 
and electrons, $T_{\rm e,max}$, radiative flux, $F_{\rm rad}$, 
integrated in the entire corona, and gas pressure, $p_{\rm tr}$, 
at location where $T=10^5$K in the TR. 
The physical interpretations of these six properties can be given as follows 
(e.g. Lamers \& Cassinelli 1999): 
$(n_{\rm p}v)_{\rm 1AU}$ is determined by the momentum and heat input 
in the subcritical region, $r<r_{\rm cr}$.  
$v_{\rm 1AU}$ reflects the momentum and heat input in the supercritical 
region. 
$T_{\rm p,max}$ and $T_{\rm e,max}$ are determined by the heating 
from the wave dissipation and also by the collisional coupling between the 
protons and electrons. 
$p_{\rm tr}$ is essentially the same as the density at the inner boundary. 
$F_{\rm rad}$ reflects the density at the coronal base, since the radiative 
flux is $\propto \rho^2$ for the optically thin plasma and the most of the 
radiation comes from the dense inner corona ($<1.5R_{\odot}$). 

\begin{figure*}
\epsfxsize=14cm
\epsfysize=13cm
\epsfbox{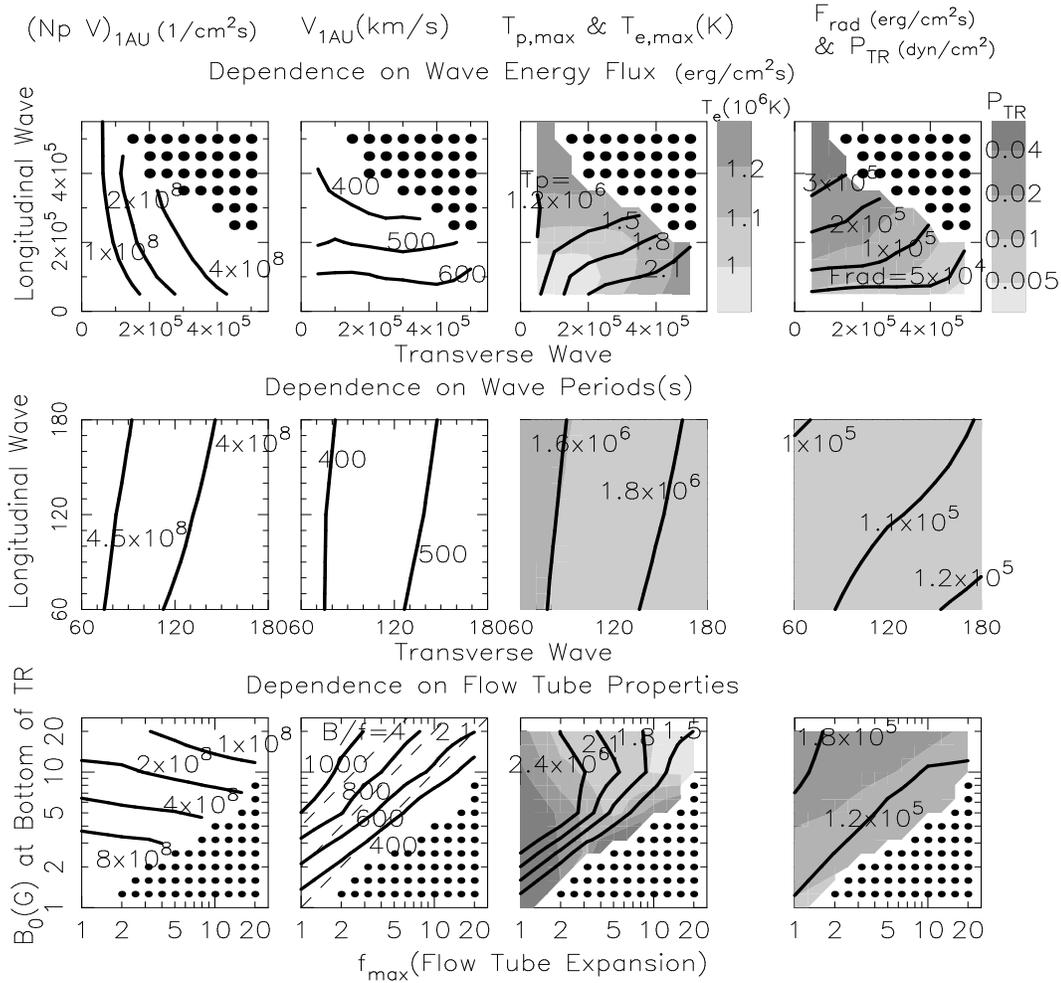}
\caption{Contours showing dependency of output properties of the solar 
corona and the solar wind on the input parameters. 
Panels on the top show dependency on $\Fwppi$ and $\Fwpri$, those 
on the middle show dependency on $\tau_{\perp}$ and $\tau_{\parallel}$, 
and those on the bottom show dependency on $f_{\rm max}$ and $B_0$. 
Panels on the first column present $(n_{\rm p}v)_{\rm 1AU}
(10^8{\rm cm^{-2}s^{-1}})$, those on the second column present 
$v_{\rm 1AU}$(km/s), those on the third column show 
$T_{\rm p,max}(10^6{\rm K})$ (solid lines) and $T_{\rm e,max}(10^6{\rm K})$ 
(filled contours in gray scale), and those on the fourth column show 
$F_{\rm rad}(10^5{\rm erg\; cm^{-2}s^{-1}})$ (solid lines) and 
$p_{\rm tr}$(dyn cm$^{-2}$) (gray contours).
Thick dots indicate the parameter sets for which we fail to integrate 
the energy equations due to too low temperature in the outer region.
Dashed lines in panel in the second column on the bottom indicate 
constant $B_0/f_{\rm max}=$4, 2, \& 1.
}
\label{fig:dpwvpr}
\end{figure*}

As for parameters, $h_{\perp}$ and $h_{\parallel}$, which control the 
energy distribution to the protons and electrons 
from the wave dissipation, we only present the results of the equipartition 
case adopting $h_{\perp}=h_{\parallel}=0.5$.
This must be reasonable for 
the N-waves, since they work only in the inner corona where the protons and 
electrons are sufficiently coupled. On the other 
hand, the approximation might be incorrect for the switch-on shock trains 
travelling in the outer region ($r>2R_{\odot}$). 
More energy is injected to the much more 
massive protons, while the lighter electrons move across the shocks  
transparently without getting so much energy. Consequently, $h_{\perp}$ 
might be smaller, though it is inevitable to perform the kinetic simulation 
to determine $h_{\perp}$ correctly. 
We have performed the calculation for several models by adopting 
the extreme value of $h_{\perp}=0$, corresponding 
to the situation that all the dissipated energy is supplied to the 
protons. Although these calculations yields larger proton temperature and 
smaller electron temperature than the equipartition cases, they give quite 
similar structures of the density and the wind velocity. 
 
Figure \ref{fig:dpwvpr} illustrates dependency of the output properties on 
the input parameters in contours. The fiducial values are set as 
$(\Fwpri, \Fwppi, \tau_{\perp}, \tau_{\parallel}, f_{\rm max}, B_0)
=(3.0\times 10^5, 2.0\times 10^5,120,120,5,5)$, where $\Fwppi$ and 
$\Fwpri$ in erg cm$^{-2}$s$^{-1}$, $\tau_{\perp}$ and $\tau_{\parallel}$ in 
second, and $B_0$ in Gauss. 
Four panels on the top show dependence on $0.5\times 10^5 \le \Fwppi, 
\Fwpri \le 5.5\times 10^5$,  
those on the middle show dependence on $60 \le \tau_{\perp}, \tau_{\parallel}
\le 180$, and those on the bottom show dependence on 
$1\le f_{\rm max}\le 20$ and $1\le B_0\le 20$. 
The first column presents contours of 
$(n_{\rm p}v)_{\rm 1AU}$, the second column presents those of $v_{\rm 1AU}$, 
the third column shows those of $T_{\rm p,max}$ and $T_{\rm e,max}$, 
and the fourth column shows those of $F_{\rm rad}$ and $p_{\rm tr}$. 
Thick dots in panels on the top and bottom denote models for which we cannot 
determine the unique coronal structure because we fail to integrate stably 
the second order derivative of the thermal conduction term due to too low 
temperature in the outer region.

Four panels on the middle indicate that dependence on the wave periods is 
quite weak within a range of $60 \le \tau_{\perp},\tau_{\parallel} \le 180$. 
In particular almost no dependence results in with respect to 
$\tau_{\parallel}$ because 
the most of the energy of the acoustic waves is lost as the downward thermal 
conduction, being unrelated to $\tau_{\parallel}$. Weak but finite 
dependences on $\tau_{\perp}$ are consistent with the results obtained by 
Nakariakov et al.(2000). 

Dependency of $(n_{\rm p}v)_{\rm 1AU}$ can be understood in terms of 
the energy and momentum deposition in the subcritical region. 
Larger input of $\Fwppi$ and $\Fwpri$ directly enhances the energy and 
momentum inputs to raise $(n_{\rm p}v)_{\rm 1AU}$ (panel in the first 
column on the top). Smaller $B_0$ makes 
the switch-on shocks dissipate more quickly, since the shock strength, 
$\alpp$, (eq.(\ref{eq:alfwvamp1})), is scaled as $\propto 
v_{\rm A}^{-3/2} \propto B_{\parallel}^{-3/2}$ (we neglect $v$ in 
eq.(\ref{eq:trwven})) for fixed $\Fwpp$. As a result, small $B_0$ increase 
the energy and momentum injection in the subcritical region to enhance    
$(n_{\rm p}v)_{\rm 1AU}$ (panel in the first column on the bottom).

A panel in the second column on the top shows that higher $v_{\rm 1AU}$ is 
achieved by models with smaller $\Fwpri$. This is because smaller input 
wave energy flux gives less dense corona and the smaller amount of coronal 
gas can be effectively accelerated to higher velocity.
One of the most important panels is that in the second column on the bottom 
which illustrates that $v_{\rm 1AU}$ is well correlated 
with $B_0/f_{\rm max}$.  $B_0/f_{\rm max}$ determines $B_{\parallel}$ 
in the outer region of $r>r_1+\sigma$ where the flow tube expands almost 
radially (eq.(\ref{eq:divb})). In models adopting 
larger $B_0/f_{\rm max}$ the acceleration and heating occurs more in the 
supercritical region than in the subcritical region because $\alpp \propto 
B_{\parallel}^{-3/2}$, and larger $v_{\rm 1AU}$ yields.
Interestingly, the nice correlation between $v_{\rm 1AU}$ and 
$B_0/f_{\rm max}$ is also obtained in recent observation \citep{hir03}, 
which might be explained by the similar discussion.

Comparison of panels in the third column with those in the first column 
show that correlation between $T_{\rm e,max}$ and $(n_{\rm p}v)_{\rm 1AU}$ is 
good. This indicates that $T_{\rm e,max}$ is mainly determined by 
the energy input in the subcritical region. On the other hand, 
$T_{\rm p,max}$ is regulated not only by the energy deposition but by the 
coupling with the electrons; too large input of energy in the subcritical 
region decreases $T_{\rm p,max}$ since more energy is transfered to the cooler 
electrons due to the larger density. 

Three panels in the fourth column illustrate that 
$F_{\rm rad}$ and $p_{\rm tr}$ are determined almost solely by $\Fwpri$. 
The N-waves rapidly dissipate and the heating occurs mainly in the inner 
corona ($\lesssim 1.5R_{\odot}$).
The dissipated energy is mostly lost as the downward 
conduction which finally radiates away in the TR 
\citep{hm82a,hm82b}. 
The radiative loss is determined by the 
density there according to $\propto \rho^2$. 
Hence, the energy from the N-wave dissipation controls the density at the 
coronal base and the TR. 
In other words, larger input of the energy in the inner 
corona can heat the plasma deeply down to the chromosphere.

Before closing this section, we summarize several important points 
with respect to the dependency on the input parameters. 
\begin{enumerate}
\item{$F_{\rm rad}$ and $p_{\rm tr}$, or equivalently the density at the 
coronal base, is mostly determined by $\Fwpri$.}
\item{The solar wind speed, $v_{\rm 1AU}$, is well correlated with 
$B_0/f_{\rm max}$, whereas it is anticorrelated with $\Fwpri$.}
\item{Within a range of $60{\rm s}\le\tau_{\perp},\tau_{\parallel}\le 
180{\rm s}$ the coronal wind structures 
are affected by $\tau_{\perp}$ and $\tau_{\parallel}$ little.} 
\item{$(n_{\rm p}v)_{\rm 1AU}$, mass flux of the solar wind, depends 
on all the parameters.}
\end{enumerate}

\section{Comparison with the Observed High/Low-Speed Solar Wind}
\label{sc:obs}
\subsection{Structure of Corona and Solar Wind}
To test reliability of our process in the real solar corona, we study 
whether our model can account for the difference between the 
high-speed solar wind from the polar coronal holes 
and the low-speed solar wind in the equatorial region 
during the low solar activity phase. 
The aim of this section is not to show 
superiority of our process to other heating and acceleration sources 
but to explore possibilities whether our 
process can become one of the reliable mechanisms.  
Recently, two or three dimensional structures have been investigated 
extensively by numerical simulations (e.g. Roussev et al.2003). 
However, the self-consistent 
treatment of the wave propagation is still difficult in such multi-dimensional 
calculations. Therefore, our one dimensional studies play a complementary 
role with them. 

As we saw in Fig.\ref{fig:dpwvpr}, the variation of the wave periods, 
$60{\rm s}\le\tau_{\perp}, \tau_{\parallel}\le 180{\rm s}$, 
affects the results little. 
Therefore, we look for the model parameters, $\Fwppi$, $\Fwpri$, 
$B_0$, and $f_{\rm max}$, for fixed 
$\tau_{\perp}=\tau_{\parallel}=120$s for simplicity.  
We summarize below how we determine the input parameters, $\Fwppi$, 
$\Fwpri$, $B_0$, and $f_{\rm max}$, which 
reproduce the data best, based on the arguments in the previous section. 

\begin{table*}[]
\caption{Models for the high/low-speed solar wind; $\Fwppi$, $\Fwpri$, and 
$F_{\rm rad}$ in ($10^5$erg cm$^{-2}$s$^{-1}$), $B_0$ in Gauss, 
$(n_{\rm p}v)_{\rm 1AU}$ in $10^8$cm$^{-2}$s$^{-1}$, $v_{\rm 1AU}$ in km/s, 
$T_{\rm p,max}$ and $T_{\rm e,max}$ in $10^6$K, and 
$p_{\rm tr}$ in $10^{-2}$dyn cm$^{-2}$. 
Note that $B_0$ is field strength at the bottom of 
the TR and $f_{\rm max}$ is also normalized there.}
\begin{tabular}{|c||c|c|c|c||c|c|c|c|c|c|}
\hline
 &$\Fwppi$ & $\Fwpri$ & $f_{\rm max}$ & $B_0$(G) & 
$(n_{\rm p}v)_{\rm 1AU}$ & $v_{\rm 1AU}$ & $T_{\rm p,max}$ 
& $T_{\rm e,max}$ & $F_{\rm rad}$ & $p_{\rm tr}$\\
\hline
\hline
High-Speed Wind& 2.4 & 0.36 & 1 & 2 & 3.9 & 881 & 2.6 & 1.2 & .42 & .38\\
\hline
Low-Speed Wind& 4.4 & 7.2 & 8 & 10 & 4.2 & 312 & 1.4 & 1.4 & 5.5 & 5.0\\
\hline
\end{tabular}
\label{tab}
\end{table*}

\begin{enumerate}
\item{The density in the inner corona is regulated 
almost solely by $\Fwpri$. 
The observed density in the inner 
corona ($\lesssim 1.2R_{\odot}$) can be explained by 
$\Fwpri\simeq 3 - 5 \times 10^4$erg cm$^{-2}$s$^{-1}$ for the less dense 
polar holes and $\Fwpri\simeq 7 - 9 \times 10^5$erg cm$^{-2}$s$^{-1}$ 
for the denser streamer region (Fig.{\ref{fig:obs}}).}
\item{Since the solar wind speed at 1AU is well correlated with 
$B_0/f_{\rm max}$, larger $B_0({\rm G})/f_{\rm max}\gtrsim 2$ is required 
for the high-speed wind ($v_{\rm 1AU}\gtrsim 700$km/s) and smaller 
$B_0({\rm G})/f_{\rm max}\simeq 1$ is favorable for the low-speed wind 
($v_{\rm 1AU}\lesssim 400$km/s).}
\item{The observed velocity profile in the polar coronal holes shows high 
outflow speed in the low corona, 
while the low-speed wind 
appeared to be accelerated gradually (bottom panel of Fig.\ref{fig:obs}). 
Jumping to the conclusion, smaller $B_0$ is required to 
reproduce larger $v$ in the inner corona. 
Since the flow tube does not completely open to $\fmax$ in the inner corona, 
smaller $B_0$ results in smaller field strength, $B_{\parallel}$, there. 
Smaller $B_{\parallel}$ anticipates larger $\alpp$ for fixed $\Fwpp$, which 
let the transverse wave dissipate more in the inner corona. 
The input of more energy in the subcritical region enhances mass flux, $\rho 
v$. On the other hand, $\rho$ in the inner corona is independently 
determined by $\Fwpri$. Consequently, a model 
adopting smaller $B_0$ with identical $\Fwppi$, $\Fwpri$, and $B_0/\fmax$ 
gives larger $v$ in the inner corona. 
}
\item{From the above restrictions of (i)-(iii), $\Fwpri$, 
$B_0$, and $f_{\rm max}$ are fixed. The last parameter, $\Fwppi$, 
is tuned to explain the density and the temperature in the intermediate 
region, $2R_{\odot} \lesssim r \lesssim 6R_{\odot}$; too much input of 
the transverse waves gives too high temperature and also too high density 
due to too large density scale height, and vice versa.} 
\end{enumerate} 
We tabulate the adopted model parameters and the results of the physical 
properties of the corona and the solar wind in Tab.\ref{tab}. 
The structures of the corona and the solar wind in both models are 
presented in Fig.\ref{fig:obs} with the observational data. 

\begin{figure}
\epsfxsize=7cm
\epsfysize=13cm
\epsfbox{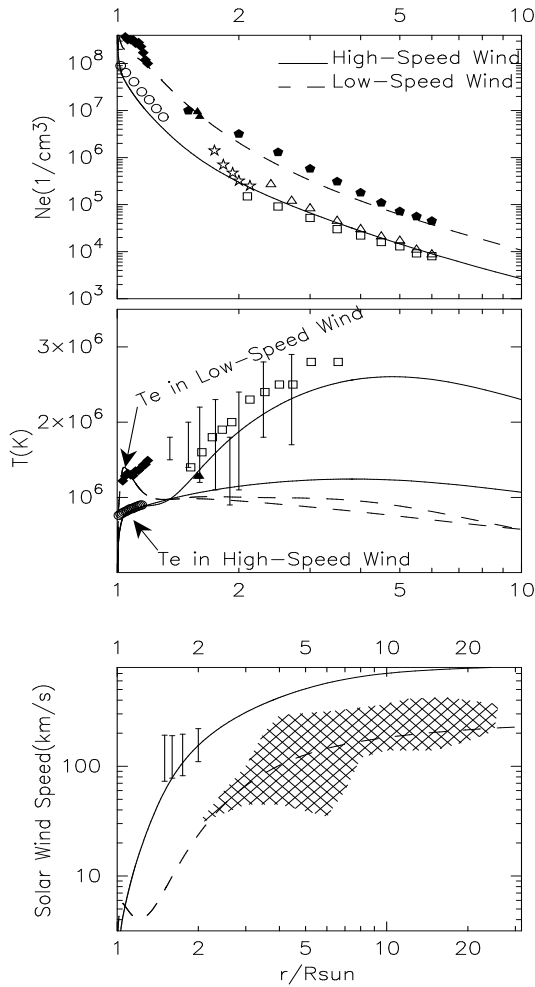}
\caption{Comparison of our results of the high-speed solar wind (solid lines) 
and the low-speed solar wind (dashed lines) with the observed data. 
In top and 
middle panels, open symbols are data in the polar coronal holes during the 
low activity phase and filled symbols are data in the equator 
or mid-latitude region where the low-speed solar wind is supposed to 
originates from.  
{\it Top} : Distribution of electron density. 
{\it Middle} : 
Distribution of proton and electron temperature. Note that proton 
temperature is higher than electron temperature in both models.
{\it bottom} : Outflow velocity of the solar wind.
Shaded region is observational data 
in the streamer belt \citep{she97}. Data with error bars 
are outflow speed based on observation of the interplume region in the 
polar holes \citep{tpr03}}
\label{fig:obs}
\end{figure}

As shown in Tab.\ref{tab} we have adopted larger $\Fwpri$ by a factor of 20 
in the low-speed wind model 
than in the high-speed wind model based on the condition (i). This 
directly leads to larger radiative flux in the low-speed wind model by a 
factor of 13, which is consistent with observational result that X-ray flux 
in the quiet-Sun in the low latitude exceed that in the polar coronal holes 
by more than an order of magnitude (e.g. \S 1 of Golub and Pasachoff 1999). 
Accordingly, the transverse waves become relatively important against the 
slow waves ($\Fwppi/\Fwpri \gg 1$) in the high-speed wind, 
while the longitudinal waves dominant ($\Fwppi/\Fwpri < 1$) in the dense 
low-speed wind. 
Table \ref{tab} also shows that the comparable order of the mass flux is 
realized in the high-speed wind model in spite of smaller input of 
$\Fwppi+\Fwpri$, 
since smaller $\fmax$ is adopted and the effect of the adiabatic expansion 
of the flow tube is reduced.

From the restrictions (ii) and (iii), $B_0$ and $f_{\rm max}$ for the 
high-speed wind model is almost uniquely determined as 
$(B_0({\rm G}),f_{\rm max})=(2,1)$.
We choose $(B_0({\rm G}),f_{\rm max})=(10,8)$ for the low-speed wind model 
which shows a little wider acceptable ranges, $6\lesssim B_0 \lesssim 12$ and 
$5 \lesssim f_{\rm max} \lesssim 10$. 
It should be noted that $B_0$, the field strength at the bottom of the TR, is 
smaller than the photospheric value which can be observed, provided that the 
field lines open out with height in the chromosphere. 
Although this ambiguity remains, our results of $B_0$ are qualitatively 
consistent with the empirical determination by \citet{sg99} giving  
surface field strength $\simeq 3$G at the pole and $\simeq 15$G 
at the equator.   
In a region of $r\gtrsim 2R_{\odot}$ the field strength 
($\propto B_0/f_{\rm max}$) in the high-speed wind is lager than that in the 
low-speed wind, which is opposed to the situation in the low corona, 
due to the flow tube expansion.

The adopted value, $f_{\rm max}=1$, for the high-speed wind model corresponds 
to the case of the radial expansion of the flow tube in the corona, 
which is consistent with the observed radial structure around $1.15 - 5.5
R_{\odot}$ in the polar coronal holes \citep{wh99}. 
However, we would like to point out that our $f_{\rm max}$ is measured from 
the bottom of the TR, which may be smaller than the observed expansion factor 
normalized by the photospheric field strength. Our results could 
become consistent with a reported expansion factor $\simeq 4$ \citep{gol96}, 
if the flux divergence of a factor of 4 occurs mainly in 
the chromosphere. 

On the other hand, our analysis has given relatively large $\fmax$ for 
the low-speed wind model, which 
means that the closed structures cover large fraction of the surface.
There have been two origins considered with respect to the low-speed wind 
(see Wang et al.2000 for review). 
One is the acceleration due to the intermittent break-up of the 
cusp-shaped closed fields on the equatorial region (e.g. Endeve, Leer, \& 
Holtzer 2003). The other is the acceleration in the open flux tube with large 
areal expansion in the low- and mid-latitude corona (e.g. Kojima et al.1999), 
which can be investigated by our modelling. \citet{kfo99} detected 
the low-speed wind with a single magnetic polarity, originating from 
the open structure region located near the active region. 
They showed the rapid expansion of the flow tube, being qualitatively 
consistent with our result,  
whereas $\fmax$ cannot be directly compared with the reported expansion factor 
$\gtrsim 50$ due to the same reason described above for the high-speed wind 
model. 

The top panel of Fig.{\ref{fig:obs}} shows electron density of the high- 
and low-speed wind model, overlayed with observational data in the polar 
region (open circles are SUMER/{\it SOHO} data by \citet{wil98}; open stars 
are UVCS/{\it SOHO} data by \citet{tpr03}; open triangles and squares are 
LASCO/{\it SOHO} data by \citet{tpr03} and \citet{lql97}) 
and in the streamer region 
(filled diamonds and triangles are CDS/{\it SOHO} and UVCS/{\it SOHO} data at  
mid-latitude by \citet{pbp00}; filled pentagons are LASCO/{\it SOHO} data 
at the equator by \citet{hvh01}).
As for the observations in the polar holes, we present data in the interplume 
regions when the plume and interplume are distinguished, according as 
\citet{tpr03} reported a strong evidence in favor of the interplume regions as 
sources of the high-speed wind.
The panel exhibits that our results can explain 
the trends of the observed electron density well in both models. 

The middle panel of Fig.{\ref{fig:obs}} compares the results of proton and 
electron temperature with the observed proton temperature 
(data with error bars are by \citet{ess99}; open squares are UVCS/{\it SOHO} 
observation by \citet{adg00}) and electron temperature (open circles are 
CDS/{\it SOHO} 
observation by \citet{fdb99}) in the polar holes and the observed electron 
temperature in the mid-latitude streamer (filled diamonds and triangles are 
CDS,UVCS/{\it SOHO} by \citet{pbp00}). Please note that the proton 
temperature is 
higher than the electron temperature in both models. 
The results of the proton and electron temperature for the high-speed solar 
wind show reasonable fits to the data. 
Our results appeared to be slightly 
lower than the observed kinetic temperature of the protons by Antonucci 
et al.(2000; open squares). 
These data indicate temperature perpendicular to the radial magnetic field. 
Although our model consider the isotropic distribution of the temperature, 
it is expected that the heating by the fast-shocks is anisotropic and 
more effective in the perpendicular direction to the field line \citep{lw00}. 
Therefore, our model could give better fitting to the data if 
taking into account the anisotropic temperature distribution. 
On the other hand, the temperature for the low-speed wind cannot reproduce 
the observed high temperature in a region of $1.3 R_{\odot}\lesssim r 
\lesssim 1.6R_{\odot}$. 
It is difficult to reproduce the observation only by adopting 
larger $\tau_{\perp}$ and $\tau_{\parallel}$ which give slower dissipation. 
Therefore, we have to say that some other heating mechanisms are required 
in this region. 

The bottom panel of Fig.{\ref{fig:obs}} presents outflow velocity of 
the solar wind.
The result of the outflow velocity for the low-speed wind is well inside of 
the obtained range by the observation of about 65 moving objects in the 
streamer.
On the other hand, our result for the high-speed solar wind is marginally 
consistent with the data of the lower edge by \citet{tpr03}. 
A cooperation with 
an extra but less dominant acceleration process would explain the data 
completely.

\subsection{Propagation of Shock Trains}
\label{sc:prwv}
Figure \ref{fig:twlw} compares the N-waves and the switch-on shock trains 
in the high- and low-speed wind models. It shows that the dissipation 
of the N-waves is much more rapid than that of the switch-on shock trains in 
both models. $\alpr$ and $\Fwpr$ is drastic decreasing functions of $r$, 
while $\alpp$ is more or less constant and $\Fwpp$ decreases gradually. 
Consequently, dissipation length, $|\mbf{\Fwpr}/\mbf{\nabla\cdot\Fwpr}|$, 
of the N-waves is very small, $\sim 0.01R_{\odot}$ in $r\lesssim 
1.5R_{\odot}$. These results 
indicate that the dissipation of the slow MHD waves is a powerful heating 
source in the inner corona, which is clearly illustrated in the top panel of 
Fig.\ref{fig:htac} comparing heating per unit mass, $Q_{\perp}/\rho$ and 
$Q_{\parallel}/\rho$. The role of the dissipation of the slow MHD waves is 
important even in the high-speed wind model, 
adopting the relatively small $\Fwpri$, to sustain the moderate density at the 
coronal base through the downward conduction.
The top panel of Fig.\ref{fig:twlw} also shows that variations of $\alpr$ 
are resemble each other in spite of the large difference of the input energy 
flux by a factor of 20. This is because the difference of the wave energy 
flux, $\Fwpr \sim \rho \alpr^2 c_{\rm s}^3$, mainly attributes to variation of 
$\rho$; larger (smaller) input of $\Fwpri$ requires larger (smaller) density 
through the energy balance between the downward thermal conduction and 
the radiative loss as discussed so far.  

\begin{figure}
\epsfxsize=7cm
\epsfysize=11cm
\epsfbox{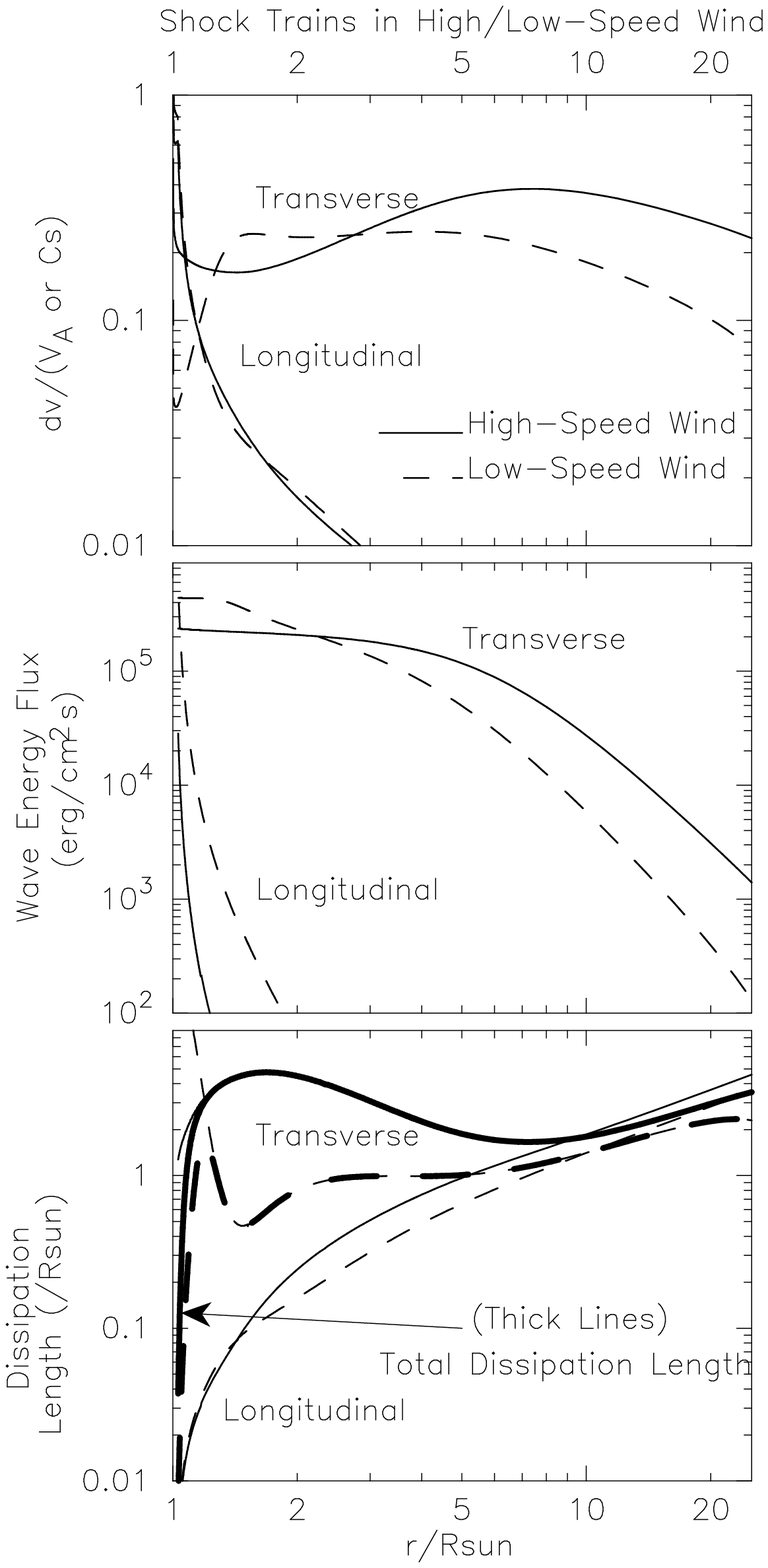}
\caption{Comparison between the switch-on shock trains and the N-waves in 
the high-speed wind (solid lines) and low-speed wind (dashed lines). 
{\it top}: Variation of shock amplitudes, $\alpp$ and $\alpr$, against radius, 
$r/R_{\odot}$. {\it middle}: Variation of wave energy flux, $\Fwpp$ 
and $\Fwpr$. {\it bottom}: Variation of dissipation length. Total dissipation 
length, $|\frac{\Fwpp+\Fwpr}{\mbf{\nabla\cdot(\Fwpp+\Fwpr)}}|$, (thick lines) 
is also displayed with dissipation length for the switch-on shock trains, 
$|\frac{\Fwpp}{\mbf{\nabla\cdot\Fwpp}}|$, and for the N-waves, 
$|\frac{\Fwpr}{\mbf{\nabla\cdot\Fwpr}}|$.}
\label{fig:twlw}
\end{figure}

\begin{figure}
\epsfxsize=7cm
\epsfysize=9cm
\epsfbox{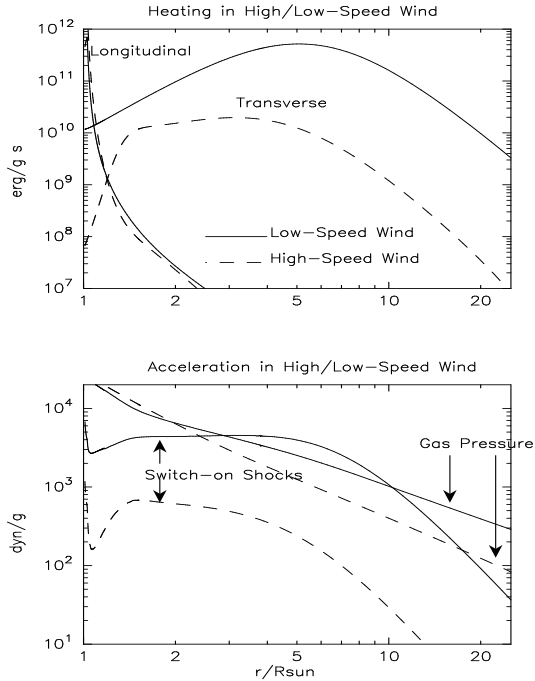}
\caption{Comparison of heating and acceleration in the high-speed wind 
(Solid lines) and low-speed wind (dashed lines). {\it top}: 
Heating per unit mass from the switch-on shock trains, $Q_{\perp}/\rho$, and 
from the N-waves, $Q_{\parallel}/\rho$. {\it bottom}: Momentum input per 
unit mass from the switch-on shock trains, $dp_{\rm w,\perp}/(\rho dr)$, is 
compared with that from the gas pressure, $dp/(\rho dr)$.
}
\label{fig:htac}
\end{figure}

The heating from the N-waves is soon dominated by the heating 
from the switch-on shocks at $\sim 1.1R_{\odot}$ in the high-speed wind model 
and $1.2R_{\odot}$ in the low-speed wind model.   
The heating from the switch-on shock trains keeps till a more distant 
region, which is seen in the top panel of Fig.\ref{fig:htac}.  
Figure \ref{fig:twlw} further shows that the dissipation of the switch-on 
shock trains 
is more rapid in the low-speed wind model, since the shock amplitude, 
$\alpp$, is larger in the inner corona 
(the top panel). This is because 
$\alpp(\propto \sqrt{\rho^{1/2} B_{\parallel}^{-3}\Fwpp})$ becomes 
systematically larger in the dense low-speed wind with smaller $B_0/\fmax$. 
   
The dissipation of the switch-on shock trains also contribute directly to 
the acceleration by the gradient of the wave pressure 
(eq.(\ref{eq.wvprss})). The bottom panel of Fig.\ref{fig:htac} presents 
the acceleration per unit mass, $dp_{\rm w,\perp}/\rho dr$, from the switch-on 
shocks compared with the acceleration from the gas pressure gradient, 
$dp/\rho dr$. Note that the acceleration from the N-waves 
is assumed to be zero as it is negligible.
The panel indicates that the 
acceleration by the switch-on shock trains dominates that by the thermal 
pressure in a region of $3-10R_{\odot}$ in the high-speed wind model, 
while the gas pressure dominates 
in the entire region in the low-speed wind model. 
In summarize, the switch-on shock trains contribute both to the heating and 
to the acceleration in the high-speed wind, while the low-wind wind 
is basically driven by the thermal pressure of the coronal 
plasma which is originally heated by the dissipations of the switch-on shock 
trains and the N-waves.   

The results shown in the bottom panel of Fig.\ref{fig:twlw} allows us to 
check the validity of the conventional assumption of the constant dissipation 
length. Total dissipation lengths, 
$|(\Fwpp+\Fwpr)/(\mbf{\nabla\cdot(\Fwpp+\Fwpr)})|$,  (thick lines) are 
at first subject to those of the N-waves in the low corona and eventually 
to those of the switch-on shock trains in the upper region ($\gtrsim 1.3 
R_{\odot}$).  
Because of the transition, the total 
dissipation lengths rapidly increase. Even in the outer region they are not 
constant; they vary $2-6R_{\odot}$ in the high-speed wind model and 
$0.4-2R_{\odot}$ in the low-speed wind model. Therefore, the assumption of 
the constant dissipation length is not adequate for our processes.

We would like to discuss the amplitudes, $\alpp$ and 
$\alpr$, in connection with the observed non-thermal broadening.
Velocity amplitude of the transverse mode in the high-speed wind model becomes 
$<\delta v_{\perp}>=\frac{1}{\sqrt{3}}\alpp v_{\rm A}\sim 
1000({\rm km/s})\times 0.15/\sqrt{3} \simeq 90$(km/s) at its maximum level, 
where a factor, $\frac{1}{\sqrt{3}}$, comes in for the trains  
illustrated in Fig.\ref{fig:wvst}. 
On the other hand, the off-limb observation in the polar coronal 
holes gives non-thermal broadening of $\delta v_{\rm obs} \simeq 30 - 50$km/s 
\citep{btd98,dtb99}, which shows that
our $<\delta v_{\perp}>$ is not consistent with $\delta v_{\rm obs}$. 
However, since uncertainties of the projection effects might affect 
$\delta v_{\rm obs}$, it would be hasty to conclude that our model 
is ruled out by the observations. 
Velocity amplitude of the longitudinal mode is as large as 
$\delta v_{\parallel}=\frac{1}{\sqrt{3}}\alpr c_{\rm s}\sim 50$km/s at 
the coronal base in both models, and it decreases to 
$\delta v_{\parallel}<10$km/s at $r=1.3R_{\odot}$. 
Nonthermal broadening of 20-40km/s is observed in the solar disk 
\citep{edpw98}.
Our results are consistent with the observed result except at the coronal 
base. 
It is also uncertain whether the discrepancy at the coronal base is real 
or not, since the same ambiguity comes into the observations of 
the longitudinal mode.

\section{Summary}
We have investigated the coronal heating and the acceleration of the high- 
and low-speed solar wind in the open-field regions by the dissipation of 
the shock trains. 
The small reconnection events as well as the surface granulations excite 
MHD waves not only at the photospheric level but in the corona. 
Among these waves, the linearly polarized \Alfven waves and acoustic waves 
steepen through the upward propagation and travel as the switch-on shock 
trains and N-waves respectively. 

Firstly, we derived evolutionary equations for the amplitudes of the shock 
trains under the WKB approximation by assuming that the dissipation occurs 
only through the shocks. 
While the assumptions are supposed to be reasonable for the longitudinal 
mode, we have to be careful in dealing with the transverse mode because 
other types of the dissipation mechanisms may be important 
as well as the WKB approximation becomes marginal in the low 
corona for the waves with periods of a few minutes. In the future studies, 
it is required to treat the transverse waves dynamically 
as taken in \citet{od97,od98,glh02}, whereas they 
consider the wave propagation under the one-fluid approximation. 
 
We determined coronal wind structure consisting of the protons and 
electrons from the bottom of the TR to 1AU by solving the wave equations 
simultaneously with the momentum and energy equations.
Studies on the dependency on the input parameters give the two 
important results: 
\begin{enumerate}
\item{$v_{\rm AU}$ has nice correlation with the field strength 
in the outer region which is proportional to $B_0/f_{\rm max}$.}
\item{The density at the coronal base is almost solely determined by $\Fwpri$.}
\end{enumerate}
In the low-speed wind, the acoustic waves play a role to keep the dense corona 
and smaller $B_0/\fmax \simeq 1$ is required to hold the wind speed slow. 
On the other hand, in the high-speed wind the transverse waves become 
relatively important and larger $B_0/\fmax \gtrsim 2$ is favorable.  
The early onset of the acceleration in the 
high-speed wind also requires the weak field strength in the low corona. 
This accordingly leads to the radial expansion of the flow tube in the 
corona, though it is not inconsistent with the observed expansion factor 
$\simeq 4$ in the polar holes if the flux divergence occurs in the 
chromosphere.

Our model has reproduced the overall trend of the high- and low-speed solar 
wind except the observed high temperature in the mid-latitude corona 
connected with the low-speed solar wind.
Therefore, our processes could be reliable heating and acceleration 
mechanisms, whereas contribution from other heating sources are required 
in the low-speed wind region.

\section*{acknowledgment}
The author thanks Drs. V. Nakariakov, K. Shibata, T. Kudoh, S. Nitta, 
S. Tsuneta, T, Sakurai, T. Terasawa, and S. Inutsuka for 
many fruitful discussions.
The author is also grateful to a referee for suitable instructions to revise 
the original draft.
The author is financially supported by the JSPS Research Fellowship for Young
Scientists, grant 4607, and by a Grant-in-Aid for the 21st Century COE 
``Center for Diversity and Universality in Physics'' at Kyoto University.

\label{lastpage}

\end{document}